\documentstyle[11pt,newpasp,twoside,epsf]{article}
\def\edcomment#1{\iffalse\marginpar{\raggedright\sl#1\/}\else\relax\fi}
\reversemarginpar

\begin{document}
\title{Distance Scale, Variable Stars and Stellar Populations in Local
Group Galaxies}
 \author{G. Clementini$^1$, L. Baldacci$^2$, A. Bragaglia$^1$, 
E. Carretta$^3$, L. Di Fabrizio$^4$, R.G. Gratton$^3$, C. Greco$^2$,
M. Gullieuszik$^5$, E.V. Held$^5$, M. Maio$^1$, M. Marconi$^6$, 
F. Matonti$^2$, Y. Momany$^5$, E. Poretti$^7$, L. Rizzi$^5$, I. 
Saviane$^8$, E. Taribello$^2$}
 \affil{$^1$INAF - Bologna Observatory, via Ranzani 1, I-40127 Bologna}
%\author{L. Baldacci, C. Greco, F. Matonti, E. Taribello}
\affil{$^2$Dept. of Astronomy, 
University of Bologna, 
via Ranzani 1, I-40127 Bologna}
%\author{E. Carretta, R.G. Gratton, E.V. Held} 
\affil{$^3$INAF - Padua Observatory, vicolo dell'Osservatorio 5, I-35122 Padova}
%\author{L. Di Fabrizio}
\affil{$^4$INAF - Centro Galileo Galilei \& Telescopio Nazionale Galileo, PO Box 565, 
38700 Santa Cruz de La Palma, Spain}
%\author{M. Gullieuszik, Y. Momany, L. Rizzi}
\affil{$^5$Dept. of Astronomy, 
 University of Padua, 
vicolo dell'Osservatorio 2, I-35122 Padova}
%\author{M. Marconi}
\affil{$^6$INAF - Capodimonte Observatory, via Moiariello 16, I-80131 Napoli}
%\author{E. Poretti}
\affil{$^7$INAF - Brera Observatory, via Bianchi 46, I-23807 Merate}
%\author{I. Saviane}
\affil{$^8$European Southern Observatory, Casilla 19001, Santiago 19, Chile}
\begin{abstract}
We present an overview of 
our study of the short period variable stars 
in the Large Magellanic Cloud, and in the dwarf galaxies Fornax, Leo\,I, 
and NGC\,6822. 
Light curves are presented for 
RR Lyrae stars, Anomalous Cepheids and, for the first time, for Dwarf 
Cepheids in 
the field and in the globular cluster \#3 of the Fornax galaxy.
\end{abstract}

\section{Introduction}
Pulsating variable stars 
play a fundamental role in establishing the astronomical distance scale, 
tracing different stellar populations, and studying
radial trends, star formation history and formation mechanisms of the host
galaxy. In recent years 
we have assembled a database on the short period variable stars 
(P$<$ 4 days) in a number of Local Group galaxies. 
Important assets of our study were (1) the collection of both photometric and
spectroscopic data with wide field imagers and large aperture
telescopes (e.g. the Wide Field Imager $-$ WFI $-$ of the ESO/MPI 2.2-m
and the ESO Very Large Telescopes $-$ VLTs)  
(2) the use of DAOPHOT and ALLFRAME   
(Stetson 1994)  
reduction packages, and (3) the detection of the variable stars with 
the Image Subtraction method 
(ISIS2.1, Alard 2000). 
Here we briefly present results for 4 of the galaxies in our sample,  
namely: the Large Magellanic Cloud (LMC), Leo\,I, NGC\,6822 and 
Fornax. A global description of the project 
can be found in Clementini (2003).
\section{The LMC: the zeropoint of the astronomical distance scale}
Being the galaxy much closer to the Milky Way, 
the LMC is the first step of the
extragalactic distance scale. 
Due to its complex stellar population
several distance indicators of
both Population I and II are present in this galaxy, thus allowing a direct 
comparison of the distance scales they provide. 
\begin{figure}
\plottwo{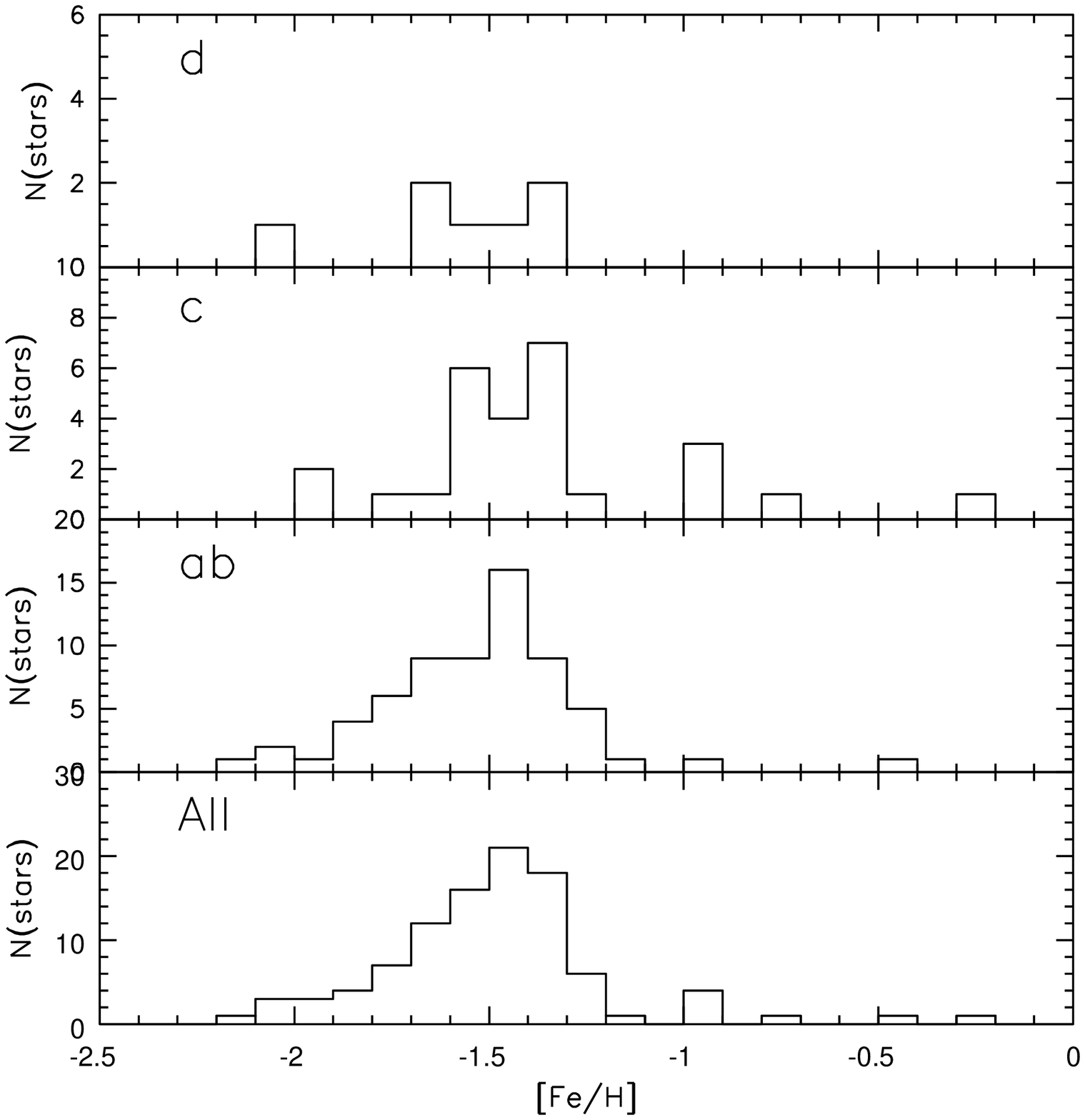}{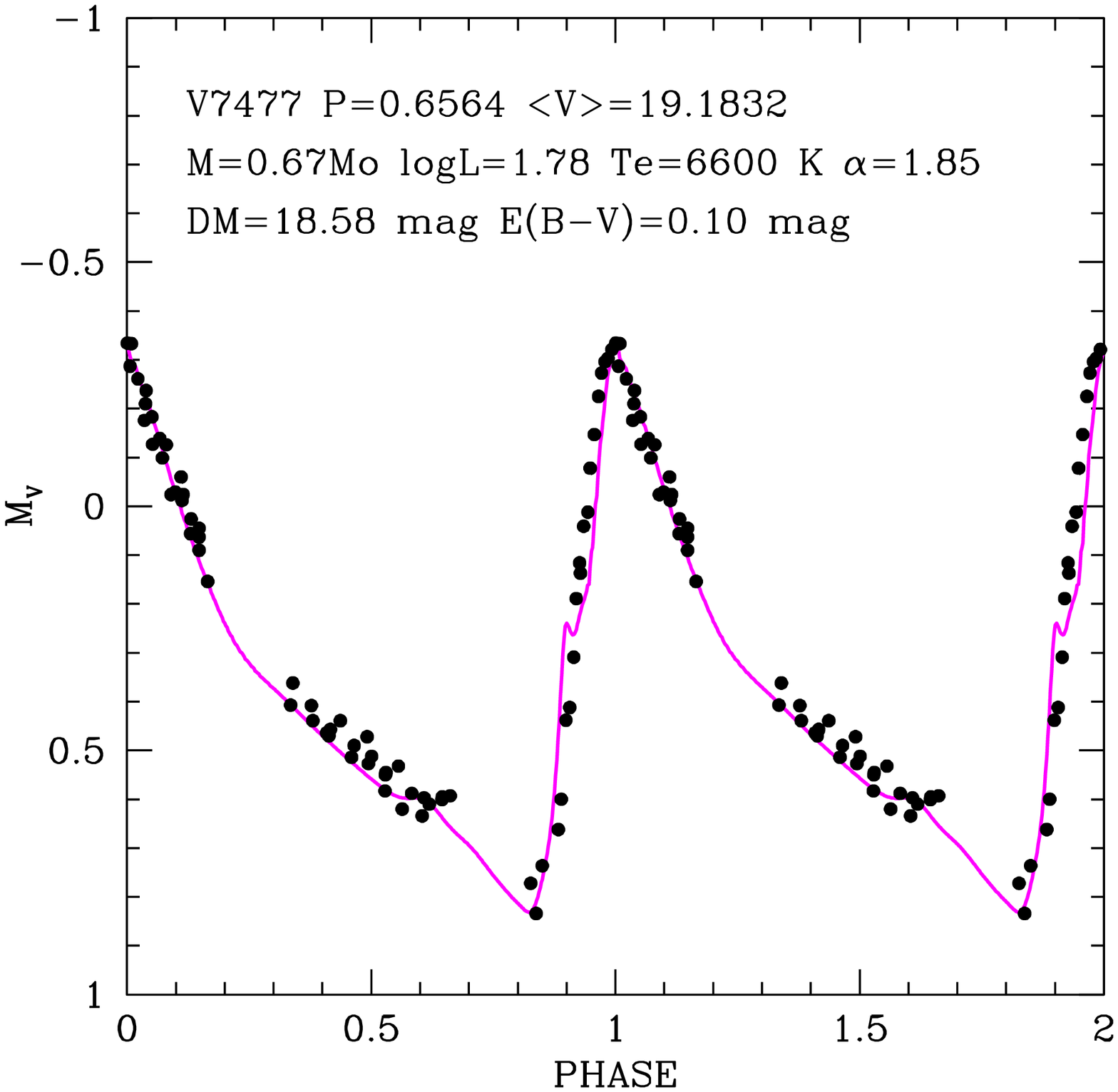}
\caption{Left panel: Metallicity distribution of
double-mode ($d$-), first overtone ($c$-), and 
fundamental mode ($ab$-type) RR 
Lyrae stars in the LMC; the bottom panel shows the total sample 
(from Gratton et al. 2003, in preparation). Right panel: Theoretical modelling 
of the $V$ light curve of the 
LMC field RR Lyrae star V7477 (from Marconi \& Clementini 2003, 
in preparation).}
\end{figure}

Our LMC variable star sample includes 
125 RR Lyrae stars (RRLs), 4 Anomalous Cepheids (ACs), 11 Cepheids, 
11 binaries, and 1
$\delta$ Scuti star 
%(see Clementini et al.\ 2003a) 
located close to the LMC bar. We have obtained $BVI$ photometry
accurate to 0.02 mag (at the luminosity level of the LMC RRLs) for the 
full sample, and individual spectroscopic metallicities 
(with FORS1 at the VLT) 
for about 80\% of the variables (Gratton et al.\ 2003, in preparation).
The metallicity distribution of the RRLs is shown in the left 
panel of Fig. 1. 
Our data allowed
(i) a very accurate definition of the average apparent magnitude of the 
LMC RRLs: $V_{0}{\rm (RR)}$=19.064$\pm$0.064, (ii) the
estimate of the local reddening from the pulsational properties of these
variables: $<E(B-V)>$=0.10$\pm 0.02$, and
(iii) the definition of the slope of the RRLs luminosity-metallicity 
relation: $\Delta M_{V}{\rm (RR)}$/$\Delta$[Fe/H]=0.214$\pm$0.047.
The corresponding distance modulus of the LMC from the average of several
independent Population I and
II distance indicators is: 
$\mu_{\rm LMC}$= 18.515 $\pm$ 0.085 mag, (Clementini et al. 2003a).
An independent estimate of 
the LMC distance was also obtained by fitting the observed RRL 
light curves with 
the theoretical pulsational models by Bono et al. (2003).
% and Marconi et al. 
%(2003, in preparation). 
An example of this theoretical modelling is shown in the right panel of Fig.
1. The distance modulus derived 
for this star is 18.58$\pm$0.10 (Marconi \& Clementini 2003, in preparation).

\section{Leo\,I and NGC\,6822: radial trends and
star formation history}
RRLs,  
ACs and Classical Cepheids allow the
unambiguous tracing of the old ($t \ga$ 10 Gyr),  
intermediate age ($t \simeq$ 1-5 Gyr) and young ($t \la$ 100 Myr) 
stellar populations across the central regions of the 
parent galaxies.
%Thanks to their luminosity variation, these stars are easily identified even 
%in the crowded central regions of the host systems.
%, where crowding hampers
%photometric resolution.

We have mapped the Leo\,I dwarf spheroidal galaxy (dSph) 
%for variable stars by
%mapping the galaxy 
with the 8 CCDs of the 
WFI mosaic. 
%at the 2.2 m ESO/MPI telescope. 
We discovered a conspicuous
population of RRLs extending all the way to the centre of the
galaxy (Held et al. 2001), and about 40 candidate ACs more concentrated 
in the inner regions of
Leo\,I (see Figure 1 of Baldacci et al. 2003a, Clementini et al.
2004, in preparation).
Our finding confirms that a first burst of stellar formation 
involving the whole galaxy 
occurred in Leo\,I about 10-13 Gyr ago, 
while the subsequent
star formation episodes that generate the candidate ACs
were more confined toward the galaxy centre.  

Time series photometry of the  
dwarf irregular galaxy NGC\,6822 was obtained with FORS2 at the VLT.
We 
%covered about $\onequarter$ of the galaxy and 
discovered about 450
candidate variables among which 17 RRLs 
%tracing the
%the first burst of star in NGC\,6822, 
and 20 brighter short-period, small amplitude variables, that we named
Low Luminosity (LL) Cepheids (Clementini et al. 2003b, Baldacci et al., 
these proceedings).
The classical instability strip in the color-magnitude diagram of
NGC\,6822 appears to be uniformly filled from the RRLs up to
the Classical Cepheids (Baldacci et al. 2003b, Held et al. 2004, 
in preparation), thus reflecting the extended 
star formation occuring in this galaxy.

\section{Fornax: galaxy formation mechanisms}
In merging scenarios the Galactic halo was assembled by accreted 
protogalactic fragments resembling the present-day  
dwarf spheroidal satellites of the Milky Way (MW).
%(Searle \& Zinn 1978, Zinn 1993,
%Mateo 1996).
A number of the Galactic globular clusters (GGCs) may originate from 
these disrupted dSphs and should keep imprinted the 
characteristics of their ``ancestors".
A marked feature of the GGCs is their division into
two sharply distinct Oosterhoff groups (Oosterhoff 1939),  
%according to the
%mean periods of their RR Lyrae stars 
however no division in Oosterhoff types is observed in the present-day dSphs 
(see Catelan, these proceedings). 
Besides Sagittarius, Fornax is the only LG dSph known to host 
globular clusters. 
We observed a WFI
%$33\times 34$ arcmin 
area of the galaxy that hosts the globular cluster Fornax\,3.
%In the preliminary analysis of the data we identified 
A sample of 108 RRLs (among which 20 double-mode candidates),
19 ACs and 25 Dwarf Cepheids (DCs) were identified in the chip of the WFI
mosaic that contains cluster \#3. Respectively 27, 8 and 2 of these variables 
belong to the cluster. 
% with WFI at the 2.2m ESO/MPI telescope.
%Variable stars were identified with ISIS2.1. Results have been obtained
%so far only for one of the 8 CCDs mosaic, the one that hosts cluster \#3
%Analysis is still in progress as preliminar results we have 
%identified 108 RR Lyrae stars (81 in the field and 27 in Fornax 3), 19 
%Anomalous Cepheids (11 field and 8 cluster's), and 25 Dwarf Cepheids 
%(23 field and 2 cluster) in this chip. 
The left panel of Fig. 2 shows the colour-magnitude diagram of
Fornax from data in this chip. The classical instability
strip is filled from the DCs up to the ACs.
%variables marked with filled symbols.
%The istability strip of Fornax is mapped out from the Dwarf Cepheidas up 
%to the anomalous Cepheids.
The right panel shows light curves of variables in Fornax\,3 (Greco 2003).
The average period of the $ab$-type RRLs is 0.595 d for the
field variables, and 0.612 d in Fornax\,3.
This $<P_{ab}>$ value places Fornax~3 at the lower edge of the 
distribution of the Oosterhoff type II GGCs.
%From these average periods we infer metallicities of [Fe/H]=$-1.77$ and $-1.91$ 
%using Sandage (1983) formulas.
%The average luminosity of the RRLs in $<V>$=21.28 and 21.21 mag 
%respectively in the field and in Fornax 3, leading to 
%distance 
%moduli of $\mu_{\rm Fornax~field}$= 20.66 $\pm$ 0.10 mag and
%$\mu_{\rm Fornax 3}$= 20.62 $\pm$ 0.11 mag (Greco 2003). 
%%stribution and thus at the limit of the cluster like the MW clusters.... 
%The average luminosity of the RR Lyrae stars in Fornax field is 21.28 and in 
%21.21. The average metallicity of the RR Lyrae stars derived from 
%the $<P_{ab}>$ and Sandage (1993) formula is: [Fe/H]=$-1.77$ and $-1.91$ in
%the field and in For3 respectively. These values lead to distance moduli of
%$\mu_{\rm Fornax field}$= 20.66 $\pm$ 0.10 mag and 
%$\mu_{\rm Fornax 3}$= 20.62 $\pm$ 0.11 mag, for $E(B-V)$=0.04 mag and
%$M_{V}(RR)$=.... (referenza).
\begin{figure}
\plottwo{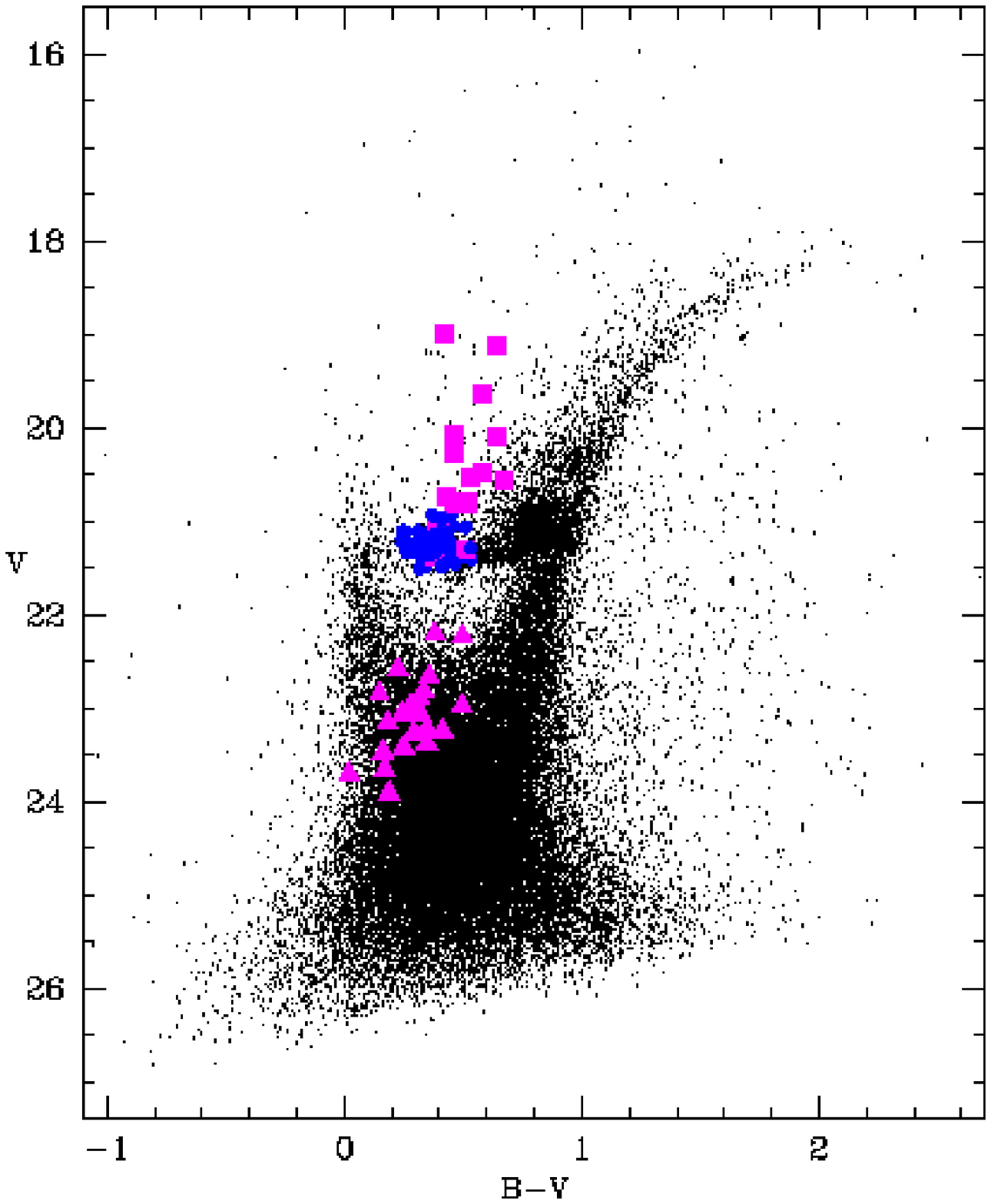}{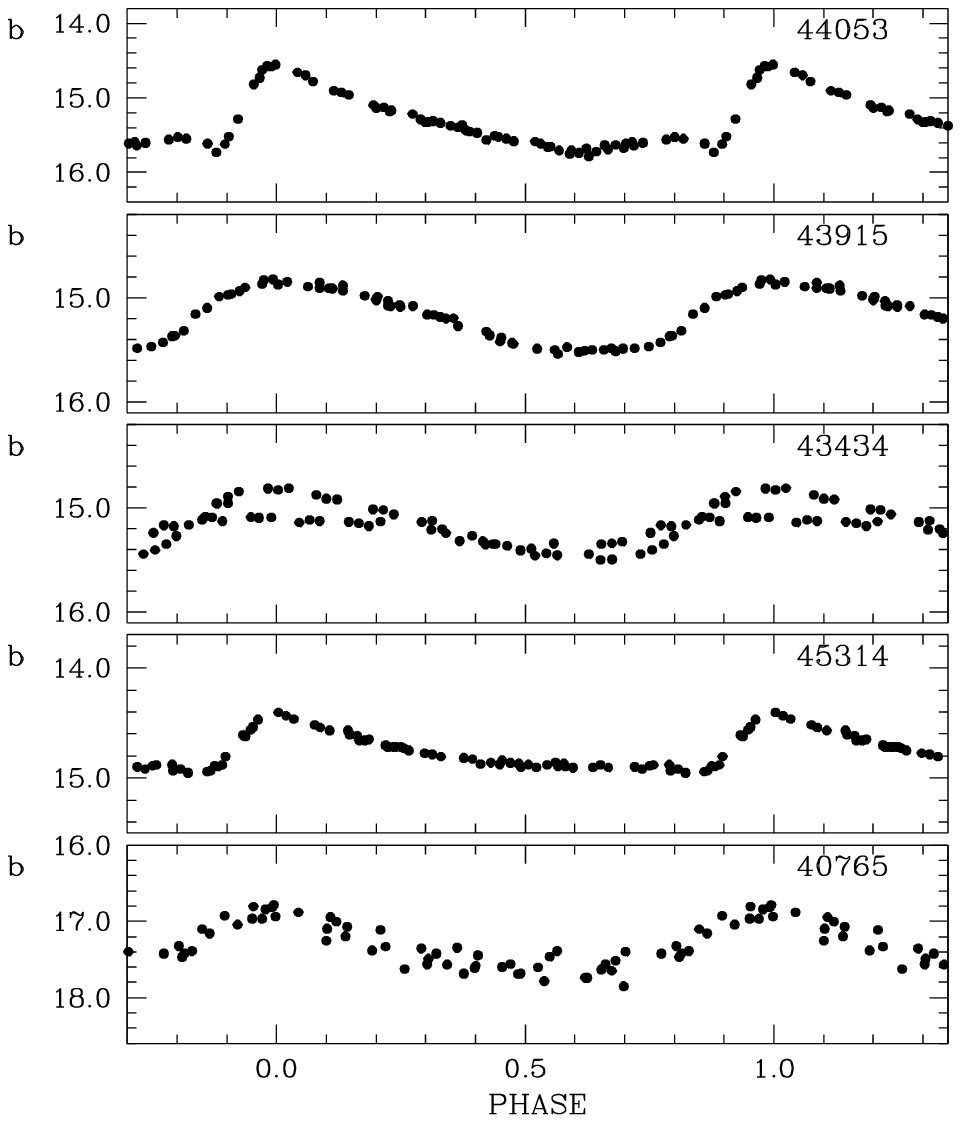}
\caption{Left panel: Colour-magnitude diagram of Fornax (from data in chip6)
with DCs, RRLs and ACs marked by
triangles, circles and squares, respectively.
 Right panel: $b$ instrumental light curves of 
$ab$-, $c$- and $d$-type RRLs, an AC, and a
DC in Fornax\,3 (from Greco 2003).} 
\end{figure}

%\section{References}
%\scriptsize

%\normalsize

\begin{references}
\reference{}Alard, C. 2000, \aaps, 144, 363
\reference{}Baldacci, L., Matonti, F., Clementini, G., Rizzi, L., Held, 
E. V., Di Fabrizio, L., Momany, Y., Saviane, I. 2003a, 
in Stars in Galaxies, ed. 
   M. Bellazzini, A. Buzzoni \& S. Cassisi, Mem.S.A.It., in press (astro-ph/0305506) 
\reference{}Baldacci, L., Clementini, G., Held, E.V., Rizzi, L. 2003b, 
   in Variability with wide-field imagers, ed. 
   G. Bono, M. Castellani \& D. Trevese, Mem.S.A.It., Vol. 74, p. 860 
   (astro-ph/0303094)
%\reference{}Baldacci, L., Rizzi, L., Clementini, G., \& Held, E.V. 2004, these 
%   proceedings
\reference{}Bono, G., Caputo, F., Castellani, V., Marconi, M., Storm, J.,
Degl'Innocenti, S. 2003, \mnras, 344, 1097
%\reference{}Catelan, M. 2004 these proceedings (astro-ph/0310159)
\reference{}Clementini, G. 2003, in Variability with wide-field imagers, ed. 
   G. Bono, M. Castellani \& D. Trevese, Mem.S.A.It., Vol. 74, p. 878 
   (astro-ph/0303067)
\reference{}Clementini, G., Gratton, R. G., Bragaglia, A., Carretta, E., 
Di Fabrizio, L., Maio, M. 2003a, \aj, 125, 1309
\reference{}Clementini, G. Held, E. V., Baldacci, L., \& Rizzi, L. 2003b, 
\apj, 588, L85
\reference{}Greco, C. 2003, Master of Science Thesis, Univ. of Bologna
\reference{}Held, E. V., Clementini, G., Rizzi, L., Momany, Y., Saviane, I., 
    Di Fabrizio, L. 2001, \apj, 562, L39
\reference{}Oosterhoff, P. Th. 1939, Observatory, 62, 104
%\reference{}Sandage, A. 1993, \aj, 106, 687
\reference{}Stetson, P. B. 1994, \pasp, 196, 250
%\reference{}Taribello, E. 2003, Master of Science Thesis, Univ. of Bologna
\end{references}
\end{document}